\documentclass{amsart}

\usepackage{amssymb,amsthm,euscript,amsmath,graphicx,array}

\def\t{& \xex{-3.9} &}                  

      \def\biT{\hskip -0.15ex}
\def\bitt{\hskip 0.30ex}     
    
\def\bitttt{\hskip 0.60ex}

\def\x{\hskip}             
\def\xex#1{\x #1 ex}

\def\szor{\biT \cdot \biT}
\def\d{{\rm d}}
\def\omit#1{}    
\def\qfc{{{\bf F}_{\rm c}}}  \def\qfd{{{\bf F}_{\rm d}}}
\def\qrhom{\tilde{\rho}}
\def\qdrho{\delta \biT \rho}
\def\qddrho{\delta' \biT \rho}

\begin{document}
\title{Stability of stationary solutions of the
    {S}chr\"odinger-{L}angevin equation}
\author{P.~V\'an* and T.~F\"ul\"op**}
\address{*Budapest University of Technology and Economics\\
Department of Chemical Physics\\
1521 Budapest, Budafoki \'ut 8.\\ and
**Institute of Particle and Nuclear Studies\\ High Energy Accelerator
Research Organization (KEK)\\ Tsukuba 305-0801, Japan}
\email{vpet@phyndi.fke.bme.hu, fulopt@post.kek.jp}

\date{\today}

\begin{abstract}
The stability properties of a class of dissipative quantum mechanical
systems are investigated. The nonlinear stability and asymptotic
stability of stationary states (with zero and nonzero dissipation
respectively) is investigated by Liapunov's direct method. The results
are demonstrated by numerical calculations on the example of the damped
harmonic oscillator.
\end{abstract}


\maketitle


\section{Introduction}

The study of dissipative quantum mechanical systems is of
widespread interest in different areas of physics
\cite{Wei93b,Men93b}. In several approaches the microscopic
systems are embedded in an environment and the dissipation is
given with a detailed description of that assumed macroscopic
background. However, the structure of the environment is
principally unobservable or unknown, moreover, its details are not
very important but only some basic properties. Several
developments of the stochastic approach to decoherence and quantum
dissipation can be interpreted as a search after these important
properties and get rid of the details of the background. The
unacceptable negative densities in the soluble Caldeira-Leggett
model \cite{CalLeg83a} can be cured by additional damping terms
\cite{Dio93a} according to the requirement of positivity of the
irreversible dynamics \cite{Lin75a}. However, complete positivity
alone does not give automatically physically meaningful results,
there are indications that some further refinements are necessary
\cite{Vac00a,FolAta01a}.

On the other hand, the approaches from the phenomenological,
thermodynamic side \cite{Ber85a,MusKau94a,KatAta00a,Kau96t} are
fighting with the right interpretation and properties of thermodynamic
quantities in quantum systems.

The situation is further complicated with the fact that there is no
common agreement in the nature of quantum dissipation, there is no
unique quantum version of the simplest dissipative classical systems
like the damped harmonic oscillator. The problem is that the lack of a
uniformly accepted variational principle prevents to use normal
canonical quantization \cite{Bol98a}. Hence quantization can be
accomplished in several different ways and the quantizations based on
different variational principles are not equivalent at all, moreover
show peculiar physical properties. The best example and a kind of
parade ground of the different methods is the mentioned quantum
harmonic oscillator with the simplest possible damping
\cite{CisLop01a,UmAta02a}.

In this paper we approach dissipation in quantum mechanics from a
different point of view and, instead of the von Neumann equation, we
investigate the properties of some simple generalizations of the
Schr\"odinger-Langevin equation, where the damping force is negatively
proportional to the velocity
\begin{equation}
i \hbar \frac{\partial \psi}{\partial t} = -\frac{\hbar^2}{2 m}
\Delta\psi + U\psi - \frac{i\hbar k}{2} \bitt
\psi\ln\frac{\psi}{\psi^*}. \label{SchroDamp1}\end{equation}

The above equation is a quantized form of Newton equation of a
classical mass-point with mass $m$, moving in a potential $U$ and with
damping term proportional to velocity
\begin{equation}
m\ddot{{\bf x}} = m \dot{{\bf v}}= -\nabla U - k {\bf v}.
\label{NewDamp1}\end{equation}

One can give several derivations of Schr\"odinger-Langevin
equation based on different assumptions
\cite{Kos72a,KanGri74a,Yas79a,Dek81a, Bol98a,Wys00a,FulKat98a} and
the equation has several appealing properties. It can be derived
by canonical quantization, the dissipation term has a clear
physical meaning, the equation preserves the uncertainty
relations, and all the stationary states of a quantized
Hamiltonian system are necessarily stationary states of the
corresponding dissipative quantum system. The later property can
be true in more general systems, too \cite{Tar02a2}. At the first
glance this last fact seems to be in contradiction with the
apparent instability of excited states \cite{Gao97a} and incited
some criticism \cite{Gre79a1} therefore it requires a more
detailed investigation that can be expected partially from
stability investigations. The relationship between the
Schr\"odinger-Langevin equation and the master equation approaches
is important also from fundamental theoretical point of view, as
it was indicated for example in \cite{CalLeg85a}.

There are some known exact solutions of equation (\ref{SchroDamp1}) for
the cases of free motion, motion in a uniform field, and for a harmonic
oscillator in one dimension \cite{Has75a,Ska75a}. These exact solutions
have the remarkable property that the stationary states of the system
are asymptotically stable.

We will show that the above properties are valid not only for these
known exact solutions of the damped harmonic oscillator but in general
for any kind of initial conditions, moreover not only for the
oscillator,  but in case of (almost) arbitrary potentials. The
stability investigation is based on the hydrodynamic model of quantum
mechanics, which recently gained a renewed interest because of the
development of powerful numerical codes of computational fluid dynamics
\cite{LopWya99a,HuAta00a,Bit00a,MadBit01a,DonMar01a}, and of its
relationship to some quantum field theories as non-Abelian fluid
dynamics \cite{BisAta02a}, in perturbational cosmological calculations
curing some instabilities of the Euler equation \cite{SzaKai02m}, etc.
In the last section the above properties will be demonstrated
numerically on the traditional parade ground of dissipative quantum
mechanics, on the example of the damped harmonic oscillator.

\section{Stability properties of the damped Schr\"odinger
equation}

Let us consider a classical mass-point with mass $m$, moving in a
potential $U$, governed by the following Newton equation:
\begin{equation}
  m\frac{\d^2{\bf x}}{\d t^2} = m\frac{\d {\bf v}}{\d t}
  = -\nabla U + \qfd,
\label{NewDamp2}\end{equation}

\noindent where $U$ is the potential of the conservative part, $\qfc
=-\nabla U$, of the total force.

It is well-known that the equation of motion of the corresponding
quantized system can be given when $\qfd={\bf 0}$ and is the
Schr\"odinger equation, (\ref{SchroDamp1}) with $k=0$. The
Schr\"odinger equation can be transformed into various thought
provoking classical forms via the Madelung transformation
$$
\psi = R e^{i S},
$$

\noindent and by introducing the so-called hydrodynamic variables,
defining the probability density $\rho$ and a velocity field ${\bf v}$
as
\begin{eqnarray}
  \rho \t := R^2 = \psi^* \psi, \\
  {\bf v} \t := \frac{\hbar}{m} \bitt \nabla S = -\frac{i \hbar}{2m}
  \bitt \nabla \ln\frac{\psi}{\psi^*}.
\end{eqnarray}

Here the star denotes the complex conjugate. With these variables the
real and imaginary parts of the Schr\"odinger equation give
\begin{eqnarray}
 \frac{\partial \rho}{\partial t} \t + \nabla \szor (\rho {\bf v})
  =0, \label{MassBal} \\
  m\frac{\partial {\bf v}}{\partial t} \t = - \nabla\left(
  m\frac{v^2}{2} + U_q + U\right), \label{Bohmeq}
\end{eqnarray}

\noindent with the condition that the vorticity of the ``fluid" is
zero:
\begin{equation}
\omega = \nabla\times{\bf v}={\bf 0}.
\end{equation}

Here
$$
U_q = -\frac{\hbar^2}{2 m} \frac{\Delta R}{R} = - \frac{\hbar^2}{4
m \rho}\left(\Delta \rho - \frac{(\nabla\rho)^2}{2\rho}\right)
$$

\noindent is sometimes called the quantum potential, and the equations
(\ref{MassBal})-(\ref{Bohmeq}) are governing equations of the fields
$(\rho, {\bf v})$. In particular, (\ref{MassBal}) is the balance of
``mass" or probability density. The second equation (\ref{Bohmeq}) is
the Newtonian equation for a point mass moving in superposed normal and
quantum potentials. The governing equations are more apparent if we
transform them into a `comoving' form:
\begin{eqnarray}
\dot{\rho} + \rho \nabla\szor{\bf v} \t = 0, \label{substMassBal} \\
m\dot{\bf v} \t = - \nabla\left(U_q + U\right). \label{substBohmEq}
\end{eqnarray}

Here the dot derivative denotes the substantial time derivative,
$\partial_t + {\bf v} \szor \nabla$. It is easy to see that the above
equations preserve the probability and vorticity. In the hydrodynamic
formalism it is apparent that any other kind of dissipative forces that
would destroy the condition of zero vorticity would destroy the
connection to the single Schr\"odinger equation and initiate coupling
to the electromagnetic field or/and the development from pure into
mixed states. It is straightforward to transform the above system into
a true hydrodynamic form, because the quantum potential can be {\em
pressurised} that is, the corresponding force density can be written as
a divergence of a quantum pressure tensor $P_q$:
$$
\rho \nabla U_q = \nabla \szor P_q,
$$

\noindent where the quantum pressure is not determined uniquely from
this condition. It can be chosen as
$$
P_{q_1} = - \frac{\hbar^2}{4 m\rho}\left(\Delta \rho \bitt I -
\frac{\nabla\rho\circ \nabla\rho}{\rho}\right),
$$
\noindent or as
$$
  P_{q_2} = - \frac{\hbar^2}{4 m\rho}\left[ (\nabla \circ \nabla ) \rho
  - \frac{\nabla\rho\circ \nabla\rho}{\rho}\right] = \frac{\hbar^2}{4
  m} ( \nabla \circ \nabla ) \ln \rho,
$$

\noindent or in some other form that differs only in a rotation of a
vector field. Here $\circ$ is the traditional notation of tensorial
product in hydrodynamics, and $I$ is the second order unit tensor. Let
us remark that thermodynamic considerations result in a unique quantum
pressure \cite{VanFul03m}. Therefore, the second equation of the above
system, (\ref{substBohmEq}), can be written in a true Cauchy form
momentum balance as well, introducing the mass density $\qrhom = m
\rho$,
  \begin{eqnarray}
  \frac{\partial \qrhom}{\partial t} + \nabla \szor (\qrhom {\bf v})
  \t =0, \label{TMassBal} \\
  \frac{ ( \partial \qrhom {\bf v} ) }{\partial t} + \nabla\left(
  \qrhom {\bf v}\circ{\bf v} + P_q\right) \t = -\qrhom \nabla U.
\end{eqnarray}

All these kinds of transformations of the Schr\"odinger equation
are well- known in the literature with different interpretations.
The Newtonian form was investigated and popularized e.g. by Bohm
and Holland \cite{Boh51b,Hol93b}. The hydrodynamic form was first
recognized by Madelung \cite{Mad26a}) and extensively investigated
by several authors, e.g. \cite{JanZie63a,JanZie64a,Wal94a}.
Moreover there is also a microscopic-stochastic background
\cite{Fen52a,Nel66a,PenCet96b,FriHau03a}, giving a reasonable
explanation of the continuum equations. Independently of the
interpretation, one can exploit the advantages of the hydrodynamic
formulation to solve the equations or investigating its
properties.

First of all let us remark that the quantum mechanical equation
(\ref{substBohmEq}) can be supplemented by forces that are not
conservative in classical mechanics but still admit a hydrodynamic
formulation:
\begin{equation} m\dot{\bf v} = -
\nabla\left(U_q + U\right) + \qfd({\bf v}).
\label{substBohmEqD}\end{equation}

The corresponding balance of momentum is
\begin{equation}
\frac{ ( \partial \qrhom {\bf v} ) }{\partial t} + \nabla\left(
  \qrhom {\bf v}\circ{\bf v} + P_q\right) = \qrhom \left( -\nabla U +
  \qfd \right) .
\label{MomBalD}\end{equation}

It is easy to check that all force fields  of the form $\qfd = -
k(S){\bf v}$ are rotation free, therefore do not violate the vorticity
conservation and the corresponding quantized form can be calculated by
canonical quantization with the help of the velocity potential $S$
\cite{KanGri74a}. The hydrodynamic transformation gives the quantized
form of the damping force expressed with the wave function as $\qfd =
-\frac{i \hbar k(S)}{2} \nabla \ln\frac{\psi}{\psi^*}$. Moreover, the
Schr\"odinger equation is transformed into the Schr\"odinger-Langevin
form (\ref{SchroDamp1}). Let us mention here that the nonlinearity of
equation (\ref{SchroDamp1}) means only a practical, not a fundamental
problem \cite{FulKat98a}.

The real stationary solutions of the above system are those where the
substantial time derivatives are zero. Thus we can eliminate the
virtual effect of the motion of the continuum, and the stationary
solution will not depend on any kind of external observers,
stationarity becomes a frame independent property. Then, the nontrivial
($\rho \neq 0$) stationary solutions ($\rho_s, {\bf v}_s$) coincide
with the stationary solutions of the Schr\"odinger equation without
damping, and in the hydrodynamic language are characterized by the
conditions,
\begin{eqnarray}
  {\bf v}_s \t = 0, \label{StacMassBal} \\
  U_q(\rho_s)+U \t = E_s = {\it const.} \label{StacMomBal}
\end{eqnarray}

For the sake of simplicity, in what follows we choose units in which
$\hbar=1$ and $m=1$ and which make all quantities dimensionless.

In the following we investigate the stability of the stationary
solutions (\ref{StacMassBal})-(\ref{StacMomBal}) of the equations
(\ref{TMassBal}) and (\ref{MomBalD}).

Let us assume that the functions $(\rho, {\bf v})$ are two times
continuously differentiable and the density $\rho_s$ exponentially
tend to zero as $|{\bf x}|$ goes to infinity
\begin{equation}
    \lim_{|{\bf x}|\rightarrow \infty}
        \rho_s(\bf x)e^{|{\bf x}|} = 0.
\label{boucon}\end{equation}

Furthermore the dissipative force obeys the following inequality
for all possible solutions of the above equations (\ref{TMassBal})
and (\ref{MomBalD})
\begin{equation}
\int \rho {\bf v} \szor \qfd \bitt \d V \leq 0.
\label{discon}\end{equation}

\noindent Here equality is valid only if $\rho {\bf v}_s = 0$.

With these conditions we show that
\begin{equation}
  L(\rho,{\bf v}) = \int \rho \left[ \frac{ v^2}{2} +
  \frac{1}{2}\left(\frac{\nabla\rho}{2\rho} \right)^2 +
  U - E_s \right] \d V
\label{Liapfun}\end{equation}

\noindent is a Liapunov functional for the stationary solution in
question.

First we show that, (\ref{Liapfun}) is decreasing along the solutions
of the differential equation. Indeed, the first derivative (variation)
of $L$ for $\begin{pmatrix} \qdrho & \delta {\bf v} \end{pmatrix}$ is
$$
  \delta L = \int \left[ \rho {\bf v} \delta{\bf v} + \left( \frac{v^2}{2}
  + U + U_q - E_s \right) \qdrho \right] \d V +
  \oint \frac{\nabla \ln \rho}{4} \bitt \qdrho \szor \d {\bf A},
$$
where the surface for the second integral is a sphere with the
radius increasing to infinity.  That term zero at the
corresponding equilibrium (stationary) solution under the
asymptotic condition (\ref{boucon}). Therefore the derivative
according to the differential equation is
\begin{equation}
  \frac{\d L}{\d t} = - \oint \left[ \rho {\bf v} \left(
  \frac{v^2}{2} + U_q + U - E_s \right) + \frac{ \nabla \biT \ln \biT
  \rho \bitttt \nabla \szor ( \rho {\bf v} ) }{4} \right] \szor
  \d{\bf A} + \int \rho {\bf v} \szor \qfd \bitt \d V \le 0,
\label{SysDer}\end{equation}

\noindent because the first surface integral vanishes according to
the asymptotic conditions, and the second integral is negative due
to (\ref{discon}).

Next, the definiteness of the second derivative of (\ref{Liapfun}) is
investigated. Evaluating the second variation between $\begin{pmatrix}
\qdrho & \delta {\bf v} \end{pmatrix}$ and $\begin{pmatrix} \qddrho &
\delta' {\bf v} \end{pmatrix}$ that yields
\begin{multline}
  \int \begin{pmatrix} \delta{\bf v} & \qdrho & \nabla\qdrho \end{pmatrix}
  \begin{pmatrix}
   \rho & {\bf v} & {\bf 0} \\
   {\bf v} &  \frac{(\nabla \rho)^2}{4\rho^3} &
        -\frac{\nabla\rho}{4\rho^2}\\
   {\bf 0} & -\frac{\nabla\rho}{4\rho^2} & \frac{1}{4\rho}
  \end{pmatrix}
  \begin{pmatrix}\delta'{\bf v} & \qddrho & \nabla\qddrho  \end{pmatrix}
  \d V.
\label{sec-der}\end{multline}

We can see that the minors of the above matrix under the integration
are
\begin{equation}
    \Delta_1 = \rho, \qquad
    \Delta_2 = \left(\frac{\nabla\rho}{2\rho}\right)^2 - {\bf v}^2, \qquad
    \Delta_3 = -\frac{{\bf v}^2}{4\rho}.
\end{equation}

The matrix in (\ref{sec-der}) is positive semidefinite in case of
stationary solutions, therefore the linearized system is marginally
stable.

Remarks:

\begin{itemize}
\item The continuity and asymptotic conditions are valid for
stationary solutions in case of several classical potentials
(harmonic, Coulomb, etc..). Moreover, one can find different
(weaker) asymptotic conditions to ensure the differentiability of
L.

\item The physical meaning of the Liapunov function
(\ref{Liapfun}) is that it is essentially the expectation value of
the energy difference of the given and the stationary states. The
decreasing property of $L$ is simply the decreasing of energy in
our dissipative system. The Liapunov function can be expressed
with the wave function as
\begin{equation}
\hat{L}(\psi)=-\int\left(\frac{\psi^*}{8\psi}(\nabla\psi)^2 +
\frac{\psi}{8\psi^*}(\nabla\psi^*)^2 + (U - E_e)\psi\psi^* \right) \d V
\end{equation}

\item According to the structure of the second derivative of $L$, given
in (\ref{sec-der}), the function space of $(\rho, {\bf v})$ functions
looks like a Soboljev space with our asymptotic conditions and norm
$$
  \|(\rho, {\bf v})\|_{\rm S}^{} = \left\{ \int \left[ \rho^2
  + {\bf v}^2 + (\nabla\rho)^2 \right] \d V \right\}^{1/2}
$$

\item If $\qfd \equiv {\bf 0}$ then the equilibrium solution is
stable with the above conditions. We can see that the dissipative,
damping force makes the stationary solutions attractive, therefore
we can expect asymptotic stability (the conditions can be rather
involved see e.g. \cite{HolAta85a}). In ordinary quantum mechanics
without damping forces the stationary solutions can be only
stable. Furthermore, the above condition can serve as a definition
of damping (dissipative) forces, i.e.\ we can say that a force is
damping (dissipative) if $\rho {\bf v}\szor \qfd \le 0$ and $\int
\rho {\bf v} \szor \qfd \bitt \d V < 0$ for any configuration
$\rho, {\bf v}$.

\item The marginal stability indicates a rich structure of
possible instabilities and therefore the need of further stability
investigations. On the other hand a rich stability structure is
expected considering the infinite number of stationary states.
Moreover, in more than one dimension  and if the domain of the
functions $(\rho, {\bf v})$ is not the whole space (e.g. in case
of a Coulomb potential), the perturbations destroying the rotation
free streams requires to consider Casimir functions for the
stability investigations \cite{HolAta85a,Arn69a}.

\item The given conditions are capable to estimate the basin of
attraction of the different quantum states. The physical role of
dissipation is to relax the process to the lowest energy element within
a basin of attraction. Moreover, the above local (!) statement
regarding the formal stability of stationary states does not exclude
the possibility that for a given initial condition the whole set of
stationary solution is the attractor of the quantum dynamics. On the
other hand transition from one basin to another possibly can be induced
by some additional interaction introduced in the system.
\end{itemize}

\section{Stability properties of the harmonic oscillator}

As a demonstration of the previous stability properties we investigate
the one dimensional dissipative harmonic oscillator, with constant
damping coefficient. In this case the hydrodynamic equations are
simplified into the following form
\begin{eqnarray}
  \frac{\partial \rho}{\partial t} \t =
  - \frac{\partial (\rho v)}{\partial x}, \label{1dMassBal} \\
  m\frac{\partial  v}{\partial t} \t =
  - \frac{\partial}{\partial x}\left[ m\frac{v^2}{2}
  - \frac{\hbar^2}{4 m \rho}\left(\frac{\partial^2\rho}{\partial x^2}
  - \frac{1}{2\rho}\left(\frac{\partial \rho}{\partial x}\right)^2\right)
  -D\frac{x^2}{2}\right]  - k v.
\label{1dBohmeq}\end{eqnarray}

Let the initial condition be a standing wave packet ($v_{\mathrm{init}}
= 0$), having the shape $\rho_ {\mathrm{init}} = \rho_{(0, 0.05)}$, and
with its maximum shifted from the center of the oscillator potential to
the left by two units. We will investigate the time development of this
initial distribution in a potential with elasticity coefficient
$D=0.02$ and different damping coefficients. In the numerical
calculations the equations have been solved with a modified leapfrog
method, adapted to the presence of the nonlinear hydrodynamic term and
for dissipative forces. The asymptotic conditions at the infinity were
considered by extrapolating the internal velocity values at the
boundaries.

\begin{figure}[ht]
\centering
\includegraphics[height=7.5cm]{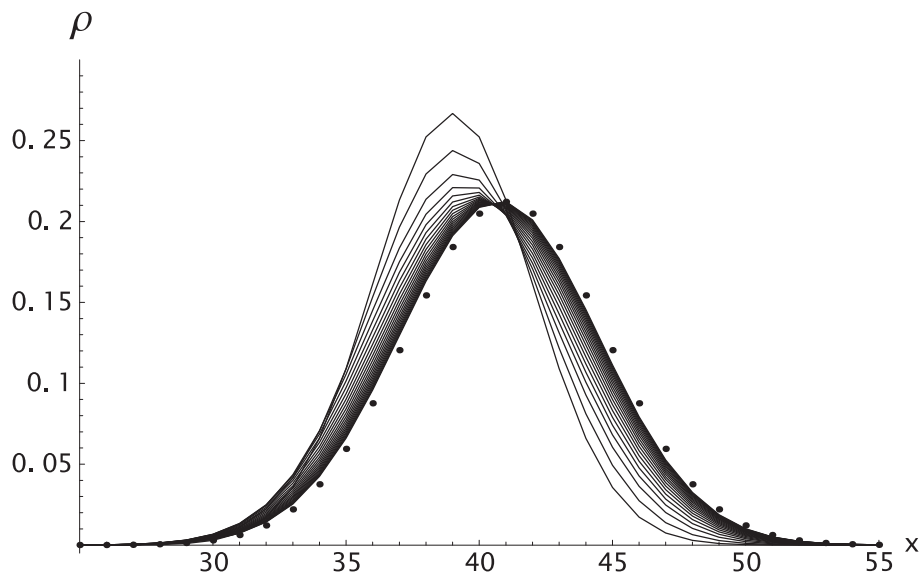}
\caption{$k=1$, large damping. Probability distribution of the
position, $t=0,..., 100$, in every 5 time units.} \label{fig1}
\end{figure}

\begin{figure}[ht]
\centering
\includegraphics[height=7.5cm]{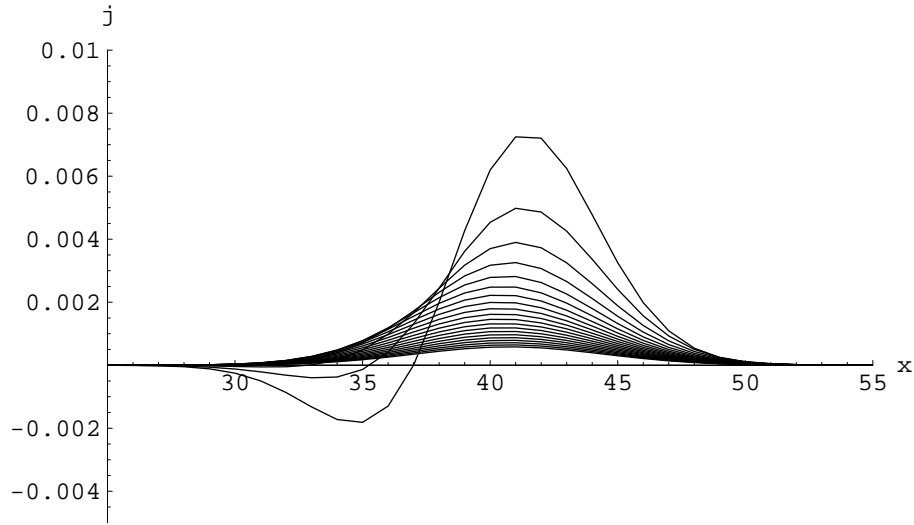}
\caption{$k=1$, large damping. Probability current of the
position, $t=0,..., 100$, in every 5 time units.} \label{fig2}
\end{figure}

On the first two figures one can see the time development of the
probability distribution and the current, respectively. In this
calculations the damping was large, with a coefficient $k=1$. On figure
\ref{fig1} the center of the oscillator is at point 41 and the
solutions starting from $\rho_{(0,0.05)}$ are given in every 5 time
units from 0 to 100. One can observe that the distribution tends toward
the stationary solution, denoted by dots. At the same time, the
probability current goes to zero as one can see on figure \ref{fig2}.
It is interesting to observe that, in the beginning, there is a space
interval where the velocity is negative and later the direction of the
motion become everywhere positive.

\begin{figure}[ht]
\centering
\includegraphics[height=7.5cm]{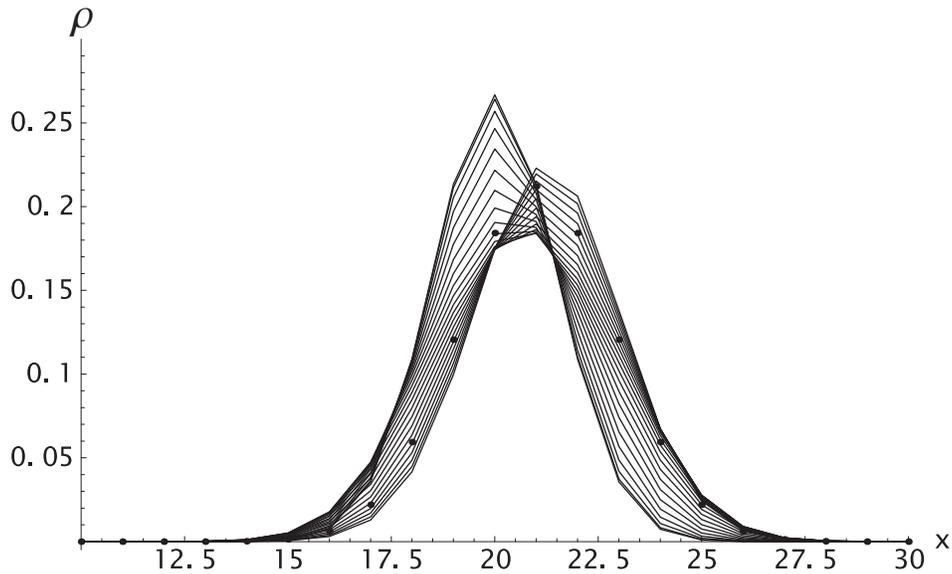}
\caption{$k=0.1$, small damping. Probability distribution of the
position, $t=0,..., 20$, in every 1 time units.} \label{fig3}
\end{figure}

\begin{figure}[ht]
\centering
\includegraphics[height=7.5cm]{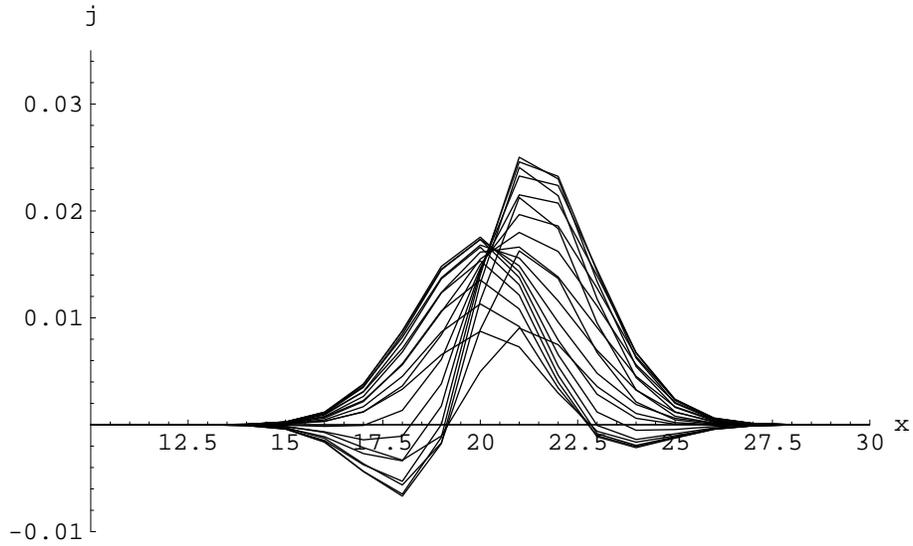}
\caption{$k=0.1$, small damping. Probability current of the
position, $t=0,..., 26$, in every 1.3 time units.} \label{fig4}
\end{figure}

\begin{figure}[ht]
\centering
\includegraphics[height=7.5cm]{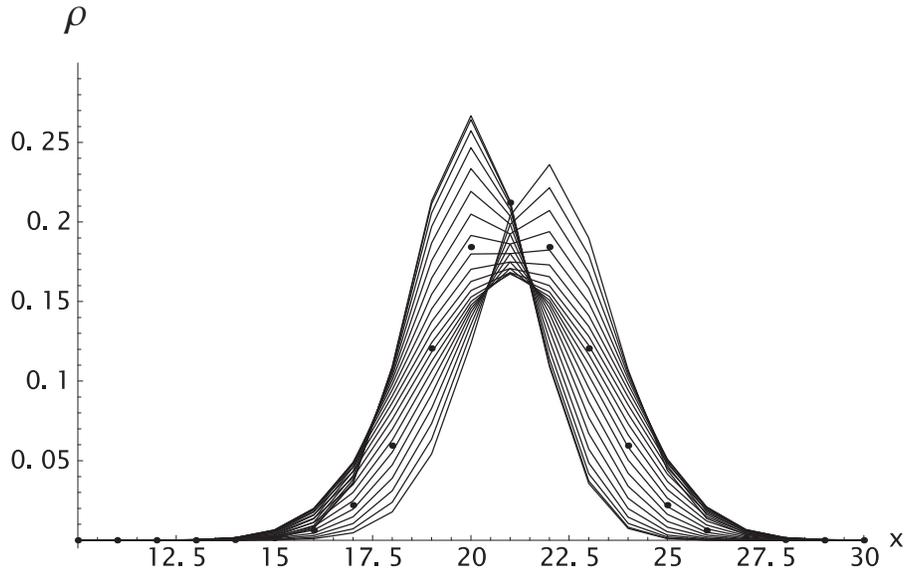}
\caption{$k=0$, zero damping. Probability distribution of the
position, $t=0,..., 19$, in every 0.95 time units.} \label{fig5}
\end{figure}

\begin{figure}[ht]
\centering
\includegraphics[height=7.5cm]{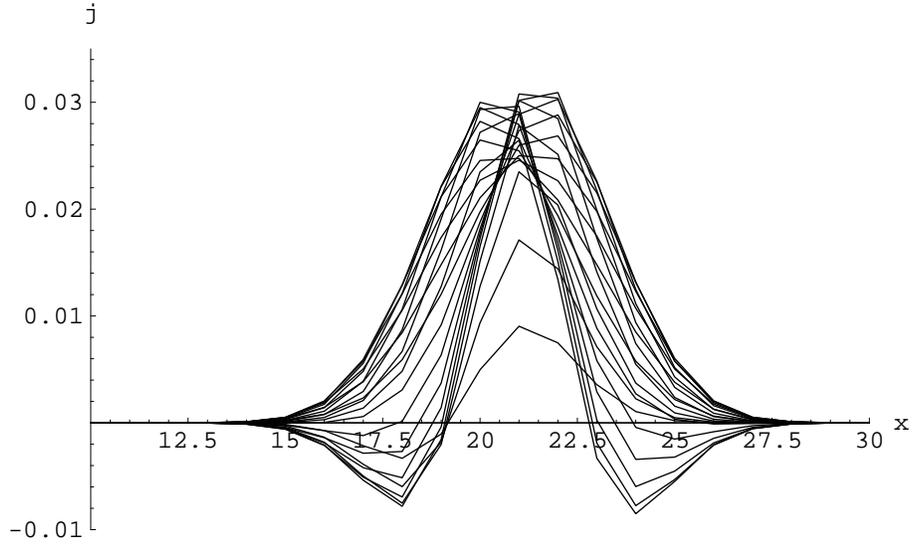}
\caption{$k=0$, zero damping. Probability current of the position,
$t=0,..., 19$, in every 0.95 time units.} \label{fig6}
\end{figure}

At the calculations leading to the next two figures the damping was
smaller, with a coefficient $k=0.1$. On figure \ref{fig3} the center of
the oscillator is at point 21 and the solutions are given in every 1
time units from 0 to 20. One can observe that the trend to equilibrium
is different from that of the previous case. A new period of a damped
oscillation starts, as one can conclude from figure \ref{fig4} of the
probability current.
At figures \ref{fig5}-\ref{fig6} there are  the undamped solutions with
$k=0$. The center of the oscillator is at point 21 and the solutions
starting from $\rho_{(0,0.05)}$ are given in every 0.95 time units from
0 to 19.

\begin{figure}[ht]
\centering
\includegraphics[height=7.5cm]{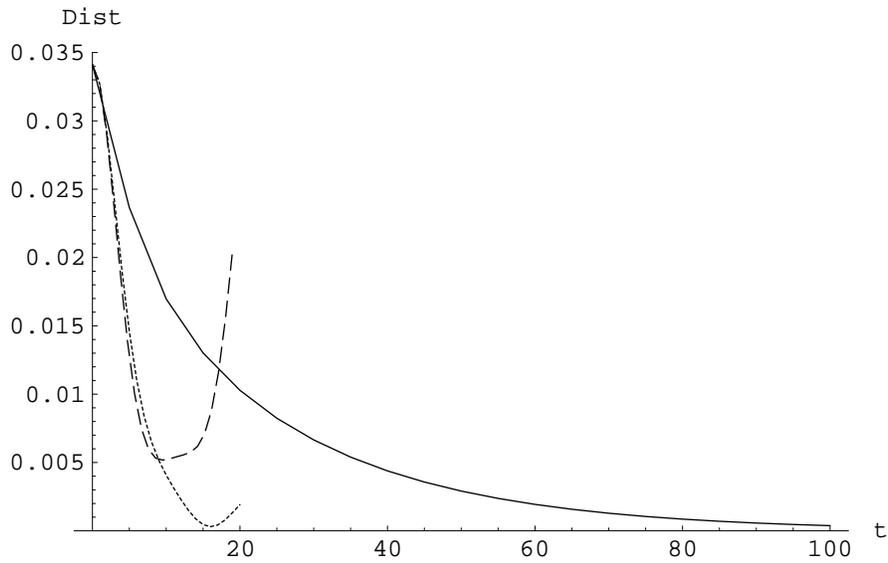}
\caption{$L^2$ distance from the equilibrium. $k=0,0.1,1$ at the
dotted, continuous and dashed lines respectively.} \label{fig7}
\end{figure}

On figure \ref{fig7} one can see the $L^2$ distance from the stationary
solution in the three cases as a function of time. Remarkably, the
starting oscillations are clearly nonlinear. The new, starting periods
indicate clearly that the $L^2$ norm is not the suitable one for
stability considerations. The expected asymptotic stability requires a
monotonous tending to the stationary solution.

\section{Conclusions}

The stability properties of the stationary solutions of the
Schr\"odinger equation were investigated. It has been shown that the
stability of stationary states can be treated by Liapunov's direct
method, where the expectation value of the energy is a good starting
point for stability investigations. However, the semidefiniteness of
the second derivative at the stationary states and the infinite number
of stationary solutions of the equation indicates a rich stability
structure. These properties are more apparent in the chosen
hydrodynamic model of quantum mechanics, enabling the use of the
methods and a direct comparison of the well understood hydrodynamic
stability problems. Moreover, the hydrodynamic analogy shows that in
one space dimension we do not need Casimirs therefore the energy is
also sufficient for exact and physically relevant results.

We have shown that in the nondissipative case, for the Shr\"odinger
equation one can expect only stability. Moreover, the above properties
were demonstrated for the damped harmonic oscillator. This property
indicates that an arbitrarily perturbed nondissipative quantum system
cannot move from a given stationary state to an other one, it will
oscillate around the perturbed state.

Finally we would like to emphasize again that the stability results are
independent on any interpretation and could be formulated with the help
of wave functions or stochastic processes, too. However, in case of
dissipation the hydrodynamic formulation has the advantage of a clear
nonequilibrium thermodynamic background with well established stability
results and clean concepts of dissipation. This kind of distinction of
dissipation types can be crucial if one would like to compare and
extend the above calculations to successful and experimentally tested
frictional models as e.g. the one of Gross and Kalinowski for heavy ion
collisions \cite{Gro75a,GroKal78a}.

\section{Acknowledgements}

This research was supported by OTKA T034715 and T034603. The
authors thank T. Matolcsi, J. Verh\'as, S. Katz, F. Bazs\'o and K.
Ol\'ah for the enlightening and interesting discussions on the
foundations and paradoxes of quantum mechanics. The calculations
have been performed by the software Mathematica. Thank for our
referee for his/her valuable remarks.

\end{document}